\input harvmac
\input epsf

\Title{\vbox{\baselineskip12pt\hbox{KOBE-TH-02-04}\hbox{{\tt hep-th/0212079}}}}
{Comments on Quiver Gauge Theories and Matrix Models}

\centerline{Shigenori Seki\footnote{$^\dagger$}{sekis@phys.sci.kobe-u.ac.jp}}
\bigskip
{\it \centerline{Department of Physics, Faculty of Science}
\centerline{Kobe University, Kobe 657-8501, Japan}}

\vskip .3in

\centerline{{\bf abstract}}

Dijkgraaf and Vafa have conjectured the correspondences 
between topological string theories, ${\cal N}=1$ gauge theories and matrix models.
By the use of this conjecture, we calculate the quantum deformations 
of Calabi-Yau threefolds with ADE singularities 
from ADE multi-matrix models. 
We obtain the effective superpotentials of the dual quiver gauge theories 
in terms of the geometric engineering for the deformed geometries.
We find the Veneziano-Yankielowicz terms in the effective superpotentials.

\Date{December 2002}

\lref\DVi{R. Dijkgraaf and C. Vafa, 
``Matrix Models, Topological Strings, and Supersymmetric Gauge Theories'', 
Nucl. Phys. B644 (2002) 3, {\tt hep-th/0206255}.}
\lref\DVii{R. Dijkgraaf and C. Vafa, 
``On Geometry and Matrix Models'', 
Nucl. Phys. B644 (2002) 21, {\tt hep-th/0207106}.}
\lref\DViii{R. Dijkgraaf and C. Vafa, 
``A Perturbative Window into Non-Perturbative Physics'', 
{\tt hep-th/0208048}.}
\lref\CM{L. Chekhov and A. Mironov, 
``Matrix Models vs. Seiberg-Witten/Whitham Theories'', 
{\tt hep-th/0209085}.}
\lref\DHKSi{N. Dorey, T. J. Hollowood, S. P. Kumar and A. Sinkovics, 
``Exact Superpotentials from Matrix Models'', 
{\tt hep-th/0209089}.}
\lref\DHKSii{N. Dorey, T. J. Hollowood, S. P. Kumar and A. Sinkovics, 
``Massive Vacua of ${\cal N}=1^*$ Theory and S-duality from Matrix Models'', 
{\tt hep-th/0209099}.}
\lref\AV{M. Aganagic and C. Vafa, 
``Perturbative Derivation of Mirror Symmetry'', 
{\tt hep-th/0209138}.}
\lref\Bo{G. Bonelli, 
``Matrix String Models for Exact (2,2) String Theories in R-R Backgrounds'', 
{\tt hep-th/0209225}.}
\lref\Fei{F. Ferrari, 
``On Exact Superpotentials in Confining Vacua'', 
{\tt hep-th/0210135}.}
\lref\FO{H. Fuji and Y. Ookouchi, 
``Comments on Effective Superpotentials via Matrix Models'', 
{\tt hep-th/0210148}.}
\lref\Be{D. Berenstein, 
``Quantum Moduli Spaces from Matrix Models'', 
{\tt hep-th/0210183}.}
\lref\DGKV{R. Dijkgraaf, S. Gukov, V. A. Kazakov and C. Vafa, 
``Perturbative Analysis of Gauged Matrix Models'', 
{\tt hep-th/0210238}.}
\lref\DHK{N. Dorey, T. J. Hollowood and S. P. Kumar, 
``S-duality of the Leigh-Strassler Deformation via Matrix Models'', 
{\tt hep-th/0210239}.}
\lref\Gor{A. Gorsky, 
``Konishi Anomaly and N=1 Effective Superpotentials from Matrix Models'', 
{\tt hep-th/0210281}.}
\lref\ACFH{R. Argurio, V. L. Campos, G. Ferretti and R. Heise, 
``Exact Superpotentials for Theories with Flavors via a Matrix Integral'', 
{\tt hep-th/0210291}.}
\lref\Mc{J. McGreevy, 
``Adding Flavor to Dijkgraaf-Vafa'', 
{\tt hep-th/0211009}.}
\lref\DGLVZ{R. Dijkgraaf, M. T. Grisaru, C. S. Lam, C. Vafa and D. Zanon,
``Perturbative Computation of Glueball Superpotentials'',
{\tt hep-th/0211017}.}
\lref\Su{H. Suzuki,
``Perturbative Derivation of Exact Superpotential for Meson Fields from Matrix Theories with One Flavour'',
{\tt hep-th/0211052}.}
\lref\Feii{F. Ferrari, 
``Quantum Parameter Space and Double Scaling Limits in ${\cal N}=1$ Super Yang-Mills Theory'', 
{\tt hep-th/0211069}.}
\lref\BR{I. Bena and R. Roiban, 
``Exact Superpotentials in $N=1$ Theories with Flavor and their Matrix Model Formulation'', 
{\tt hep-th/0211075}.}
\lref\DJ{Y. Demasure and R. A. Janik, 
``Effective Matter Superpotentials from Wishart Random Matrices'', 
{\tt hep-th/0211082}.}
\lref\AKMV{M. Aganagic, A. Klemm, M. Marino and C. Vafa, 
``Matrix Model as a Mirror of Chern-Simons Theory'', 
{\tt hep-th/0211098}.}
\lref\Gop{R. Gopakumar, 
``${\cal N}=1$ Theories and a Geometric Master Field'', 
{\tt hep-th/0211100}.}
\lref\NSWi{S. Naculich, H. Schnitzer and N. Wyllard, 
``The ${\cal N}=2$ $U(N)$ Gauge Theory Prepotential and Periods from a Perturbative Matrix Model Calculation'', 
{\tt hep-th/0211123}.}
\lref\CDSW{F. Cachazo, M. R. Douglas, N. Seiberg and E. Witten,
``Chiral Rings and Anomalies in Supersymmetric Gauge Theory'',
{\tt hep-th/0211170}.}
\lref\Ta{Y. Tachikawa,
``Derivation of the Konishi anomaly relation from Dijkgraaf-Vafa with (Bi-)fundamental matter'',
{\tt hep-th/0211189}.}
\lref\DNV{R. Dijkgraaf, A. Neitzke and C. Vafa,
``Large N Strong Coupling Dynamics in Non-Supersymmetric Orbifold Field Theories'',
{\tt hep-th/0211194}.}
\lref\Feng{B. Feng,
``Seiberg Duality in Matrix Model'',
{\tt hep-th/0211202}.}
\lref\KMT{A. Klemm, M. Marino and S. Theisen,
``Gravitational Corrections in Supersymmetric Gauge Theory and Matrix Models'',
{\tt hep-th/0211216}.}
\lref\FH{B. Feng and Y.-H. He,
``Seiberg Duality in Matrix Models II'',
{\tt hep-th/0211234}.}
\lref\KMar{V. A. Kazakov and A. Marshakov,
``Complex Curve of the Two Matrix Model and its Tau-function'',
{\tt hep-th/0211236}.}
\lref\DST{R. Dijkgraaf, A. Sinkovics and M. Temurhan,
``Matrix Models and Gravitational Corrections'',
{\tt hep-th/0211241}.}
\lref\IMi{H. Itoyama and A. Morozov,
``The Dijkgraaf-Vafa Prepotential in the Context of General Seiberg-Witten Theory'',
{\tt hep-th/0211245}.}
\lref\ACFHii{R. Argurio, V. L. Campos, G. Ferretti and R. Heise,
``Baryonic Corrections to Superpotentials from Perturbation Theory'',
{\tt hep-th/0211249}.}
\lref\NSWii{S. Naculich, H. Schnitzer and N. Wyllard,
``Matrix Model Approach to the ${\cal N}=2$ $U(N)$ Gauge Theory with Matter in the Fundamental Representation'',
{\tt hep-th/0211254}.}
\lref\IMii{H. Itoyama and A. Morozov,
``Experiments with the WDVV Equations for the Gluino-condensate Prepotential: the Cubic (two-cut) Case'',
{\tt hep-th/0211259}.}
\lref\INO{H. Ita, H. Nieder and Y. Oz,
``Perturbative Computation of Glueball Superpotentials for SO(N) and USp(N)'',
{\tt hep-th/0211261}.}
\lref\BRT{I. Bena, R. Roiban and R. Tatar,
``Baryons, Boundaries and Matrix Models'',
{\tt hep-th/0211271}.}
\lref\Oo{Y. Ookouchi,
``${\cal N}=1$ Gauge Theory with Flavor from Fluxes'',
{\tt hep-th/0211287}.}
\lref\ACHKR{S. K. Ashok, R. Corrado, N. Halmagyi, K. D. Kennaway and C. Romelsberger, 
``Unoriented Strings, Loop Equations, and ${\cal N}=1$ Superpotentials from Matrix Models'',
{\tt hep-th/0211291}.}
\lref\Ohta{K. Ohta,
``Exact Mesonic Vacua from Matrix Models'',
{\tt hep-th/0212025}.}
\lref\IMiii{H. Itoyama and A. Morozov,
``Calculating Gluino-Condensate Prepotential'',
{\tt hep-th/0212032}.}
\lref\JO{R. A. Janik and N. A. Obers, 
``$SO(N)$ Superpotential, Seiberg-Witten Curves and Loop Equations'',
{\tt hep-th/0212069}.}
\lref\Ma{J. Maldacena,
``The Large $N$ Limit of Superconformal Field Theories and Supergravity'',
Adv. Theor. Math. Phys. 2 (1998) 231; Int. J. Theor. Phys. 38 (1999) 1113,
{\tt hep-th/9711200}.}
\lref\GV{R. Gopakumar and C. Vafa,
``On the Gauge Theory/Geometry Correspondence'',
Adv. Theor. Math. Phys. 3 (1999) 1415,
{\tt hep-th/9811131}.}
\lref\Va{C. Vafa,
``Superstrings and Topological Strings at Large $N$'',
J. Math. Phys. 42 (2001) 2798, 
{\tt hep-th/0008142}.}
\lref\CIVi{F. Cachazo, K. Intriligator and C. Vafa,
``A Large $N$ Duality via a Geometric Transition'',
Nucl. Phys. B603 (2001) 3,
{\tt hep-th/0103067}.}
\lref\CIVii{F. Cachazo, K. Intriligator and C. Vafa,
``Geometric Transitions and $N=1$ Quiver Theories'',
{\tt hep-th/0108120}.}
\lref\CFIKV{F. Cachazo, B. Fiol, K. Intriligator, S. Katz and C. Vafa, 
``A Geometric Unification of Dualities'',
Nucl. Phys. B628 (2002) 3,
{\tt hep-th/0110028}.}
\lref\OT{K. Oh and R. Tatar, 
``Duality and Confinement in ${\cal N}=1$ Supersymmetric Theories from Geometric Transitions'',
Adv. Theor. Math. Phys. 6 (2002) 141,
{\tt hep-th/0112040}.}
\lref\KMMMP{S. Kharchev, A. Marshakov, A. Mironov, A. Morozov and S. Pakuliak,
``Conformal Matrix Models as an Alternative to Conventional Multi-Matrix Models'',
Nucl. Phys. B404 (1993) 717,
{\tt hep-th/9208044}.}
\lref\Ko{I. Kostov,
``Gauge Invariant Matrix Model for the \^A-\^D-\^E Closed Strings'',
Phys. Lett. B297 (1992) 74,
{\tt hep-th/9208053}.}
\lref\GM{P. Ginsparg and G. Moore,
``Lectures on 2D Gravity and 2D String Theory'',
TASI 1992, {\tt hep-th/9304011}.}
\lref\FGZ{P. Di Francesco, P. Ginsparg and J. Zinn-Justin,
``2D Gravity and Random Matrices'',
Phys. Rept. 254 (1995) 1,
{\tt hep-th/9306153}.}
\lref\KM{S. Katz and D. R. Morrison,
``Gorenstein Threefold Singularities with Small Resolutions via Invariant Theory for Weyl Groups'',
J. Alg. Geom. 1 (1992) 449,
{\tt alg-geom/9202002}.}
\lref\VY{G. Veneziano and S. Yankielowicz,
``An Effective Lagrangian for the Pure ${\cal N}=1$ Supersymmetric Yang-Mills Theory'',
Phys. Lett. B113 (1982) 231.}

\newsec{Introduction}

String theories give us a lot of useful methods 
in order for us to understand various gauge theories. 
For example, the AdS/CFT correspondence \Ma\ and 
the gauge/gravity correspondence \GV\ are known well.
A-model topological string theories correspond to Chern-Simons gauge theories 
by the gauge/gravity correspondence.

There are mirror symmetry between the A-model topological string theories 
and the B-model topological string theories.
Recently Dijkgraaf and Vafa have proposed the correspondences 
between the B-model topological string theories, 
${\cal N}=1$ supersymmetric gauge theories 
and large $N$ matrix models \refs{\DVi\DVii{--}\DViii}.
In other words, these correspondences are the mirror dual of 
the gauge/gravity correspondence.
The ${\cal N}=1$ gauge theory is constructed 
by adding certain superpotential to the ${\cal N}=2$ gauge theory.
In the description of D-brane configuration, 
the ${\cal N}=1$ gauge theory is realized 
by D5-branes wrapped on two-cycles in Calabi-Yau manifolds.
When the two-cycles are blown down, new three-cycles emerge. 
Three-form RR and NSNS fluxes then appear instead of the D5-branes.
This is called geometric transition \refs{\Va\CIVi\CIVii\CFIKV{--}\OT}.
On the Calabi-Yau manifolds after the geometric transition, 
there are two kinds of three cycles, 
which are compact $A_i$-cycles and non-compact $B_i$-cycles.
We define periods,
\eqn\introi{
\mu_i = {1 \over 2\pi i}\oint_{A_i} \Omega,\quad 
\Pi_i = {\partial {\cal F}_0 \over \partial \mu_i} = \int_{B_i} \Omega,
}
where $\Omega$ is a holomorphic three-form on the Calabi-Yau threefold
and ${\cal F}_0$ is a prepotential.
In terms of these periods we can write down 
the effective superpotential of the dual gauge theory,
\eqn\introii{
W_{\rm eff} = \sum_i (N_i\Pi_i - 2\pi i \tau_i \mu_i),
}
where $N_i$ is a number of D-branes and $\tau_i$ is a gauge coupling.

It is proposed that the effective superpotential can be reproduced 
by the matrix model with certain tree level superpotential $W(\Phi)$ \DVi.
The partition function of the matrix model is
$Z= \int d\Phi \exp\left(-{1 \over g_s}W(\Phi)\right)$.
Fixing $S_i=N_i g_s$, we take the limit $N_i \gg 1, g_s \ll 1$, 
and the partition function then leads 
to $Z = \exp \sum_g g_s^{2g - 2} {\cal F}_g(S_i)$.
The free energies ${\cal F}_g$ are the contributions of genus $g$ diagrams.
In particular ${\cal F}_0$ is derived from the planar diagrams.
If we can calculate the partition function, 
we obtain the free energy, $\log Z$.
We can identify it with the prepotential of the dual gauge theory 
and we obtain the exact effective superpotential in terms of 
\introi\ and \introii.
Since the perturbative analyses in the matrix models lead to the exact results 
in the dual gauge theories, this new derivation 
of the superpotentials is powerful.
A lot of works on this subject have been done \refs{\CM\DHKSi\DHKSii
\AV\Bo\Fei\FO\Be\DGKV\DHK\Gor\ACFH\Mc\DGLVZ\Su\Feii\BR\DJ\AKMV\Gop\NSWi
\CDSW\Ta\DNV\Feng\KMT\FH\KMar\DST\IMi\ACFHii\NSWii\IMii\INO\BRT\Oo\ACHKR
\Ohta\IMiii{--}\JO}.

In this paper we consider ${\cal N}=1$ quiver gauge theories
and matrix models. 
The matrix models are ADE multi-matrix models, which have been studied in
\refs{\KMMMP,\Ko}.
The quiver gauge theories are realized 
by the string theories on the Calabi-Yau manifolds with ADE singularities.
The ${\cal N}=2$ quiver gauge theories lead to the ${\cal N}=1$ theories 
by the additional superpotentials, while the dual Calabi-Yau geometries 
are deformed. These deformations are reproduced by the matrix model \DVii.
Since we systematically introduce a lot of gauge symmetries to 
the quiver gauge theories, they are interesting 
also for realistic particle theories \Mc.

In Section 2, we will analyse the quantum deformations of 
the ADE singularities in the matrix model side.
In terms of the deformed geometries, we will calculate 
the superpotentials of the dual quiver gauge theories.
Section 3 is devoted to the conclusions and the some comments on 
left problems.

\newsec{Effective superpotentials of quiver gauge theories}

Before discussing on the quiver gauge theories and the multi-matrix models,
we will give a brief review on a one-matrix model \DVi.
Let us consider an $N \times N$ Hermitean matrix $\Phi$.
The partition function of the one-matrix model 
with the tree level superpotential $W(\Phi)$ is
\eqn\omi{
Z = \int d\Phi \exp \left( -{1 \over g_s} W(\Phi) \right) ,
}
where we set that $W(\Phi)$ is a degree $n$ polynomial of $\Phi$.
In terms of the $N$ eigen values $\lambda_I$ $(I=1,\cdots,N)$ of $\Phi$, 
we can rewrite the partition function \omi\ as
\eqn\omii{
Z = \int \left(\prod_{I=1}^N d\lambda_I \right) 
\Delta(\lambda)^2
\exp \left( -{1 \over g_s} \sum_{I=1}^N W(\lambda_I) \right), 
}
where $\Delta(\lambda)$ is the Vandermonde determinant, 
$\prod_{I<J} (\lambda_I - \lambda_J)$.
When we describe the partition function as $Z = \int \Pi_I d\lambda_I e^{-\hat{S}}$, 
the effective action $\hat{S}$ is denoted by
\eqn\omiii{
\hat{S} = {1 \over g_s} \sum_{I=1}^N W(\lambda_I) - 2 \sum_{I<J}\log (\lambda_I - \lambda_J).
}
From the action \omiii,
the equations of motion for $\lambda_I$ are written down as
\eqn\omiv{
{1 \over g_s} W'(\lambda_I) 
- 2 \sum_{J \neq I} {1 \over \lambda_I - \lambda_J} = 0.
}
We introduce a resolvent,
\eqn\omv{
\omega(x) \equiv {1 \over N} \sum_{I=1}^N {1 \over x - \lambda_I},
} 
which is useful in the matrix model technology \refs{\GM,\FGZ}.
Physical meaning of the resolvent is a loop operator and 
we can easily derive a loop equation from \omiv\ in terms of the resolvent.
From \omiv, we define the function,
\eqn\omvi{
y(x) = W'(x) - 2g_s \sum_{J=1}^N {1 \over x - \lambda_J}
= W'(x) - 2 S \omega(x),
}
where $S$ is the 't Hooft coupling $N g_s$. 
$\omega(x)$ is not a polynomial, but, in large $N$ limit, 
$y(x)^2$ is given by $y(x)^2 = W'(x)^2 + f_{n-1}(x)$, 
where $f_{n-1}(x)$ is a degree $n-1$
polynomial.

In the context of large $N$ duality and geometric transitions 
\refs{\Va\CIVi\CIVii{--}\CFIKV}, 
the dual Calabi-Yau geometry after the deformation is denoted by
\eqn\omvii{
u^2 + v^2 + y^2 + W'(x)^2 +f_{n-1}(x) = 0.
}
We then consider the one-form
\eqn\omviii{
y(x)dx = \sqrt{W'(x)^2 +f_{n-1}(x)} dx.
}
The periods \introi\ are described as
\eqn\omix{
{1 \over 2\pi i}\int_{A_i} y(x) dx = \mu_i,\quad 
\int_{B_i}^{\Lambda^{3 \over 2}} y(x) dx 
= {\partial {\cal F}_0 \over \partial \mu_i} = \Pi_i,
}
in terms of the one-form \omviii.
Without the deformation, $y(x)$ in \omviii\ is equal to $W'(x)$.
Since the function \omvi\ derived from the matrix model 
can be identified with $y(x)$ in \omviii, 
$-2S\omega(x)$ in \omvi\ leads to $f_{n-1}(x)$ in \omvii, 
in other words, $f_{n-1}(x)$ is regarded 
as the contribution of loop operators in 
the matrix model. Adding $f_{n-1}$ is called a quantum deformation.

Let us now consider the ADE singularities and the quiver gauge theories.
The Calabi-Yau threefolds with the ADE singularities 
are realized by the fibration of two dimensional ADE singularities.
The fibres over $x$-plane are denoted by
\eqnn\fibrea
\eqnn\fibred
\eqnn\fibreei
\eqnn\fibreeii
\eqnn\fibreeiii
$$\eqalignno{
A_n \quad &u^2 + v^2 + \prod_{i=1}^{n+1} (y - t_i(x)) = 0,\quad 
\sum_{i=1}^{n+1} t_i = 0, &\fibrea \cr
D_n \quad &u^2 + v^2 y + {1 \over y}\left(\prod_{i=1}^n(y - t_i(x)^2)
- \prod_{i=1}^n t_i(x)^2\right) + 2\prod_{i=1}^n v t_i(x) = 0, &\fibred \cr
E_6 \quad &u^2 + v^3 + y^4 + \epsilon_2(x)vy^2 
+ \epsilon_5(x)vy + \epsilon_6(x)y^2 + \epsilon_8(x)v \cr
&+ \epsilon_9(x)y + \epsilon_{12}(x) = 0, &\fibreei\cr
E_7 \quad &u^2 + v^3 + vy^3 + \epsilon_2(x)v^2y + \epsilon_6(x) v^2
+ \epsilon_8(x)vy + \epsilon_{10}(x)y^2 \cr
&+ \epsilon_{12}(x)v + \epsilon_{14}(x)y + \epsilon_{18}(x) = 0, &\fibreeii \cr
E_8 \quad &u^2 + v^3 + y^5 + \epsilon_2(x)vy^3 + \epsilon_8(x)vy^2 
+ \epsilon_{12}(x)y^3 + \epsilon_{14}(x)vy \cr
&+ \epsilon_{18}(x)y^2 + \epsilon_{20}(x)v + \epsilon_{24}(x)y + \epsilon_{30}(x) = 0, &\fibreeiii
}$$
where $\epsilon_i$ are functions of $t_i(x)$ and 
are explicitly written down in \KM.
For these fibrations we can describe the one-forms $y_i(x)dx$ \CIVii\ as
\eqnn\gofan
\eqnn\gofdn
\eqnn\gofen
$$\eqalignno{
A_n\quad &y_i = t_{i+1} - t_i,\quad i=1,\cdots,n, &\gofan \cr
D_n\quad &y_i = t_{i+1} - t_i,\quad i = 1,\cdots,n-1,\qquad y_n=-t_{n-1} - t_n, &\gofdn \cr
E_n\quad &y_i=t_{i+1} - t_i,\quad i=1,\cdots,n-1,\qquad y_n=t_1 + t_2 + t_3. &\gofen
}$$
If we calculate the periods \omix\ for the one-forms \gofan, \gofdn\ and \gofen, 
we obtain the effective superpotentials of the quiver gauge theories.

In the following, we will derive the quantum deformations 
of the ADE singularities from the ADE matrix models and
consider the effective superpotentials of the dual quiver gauge theories.

\subsec{$A_n$ singularity}

Firstly we consider the $A_n$ singularities 
and the $A_n$ quiver gauge theories. 

\bigskip
\centerline{\epsfbox{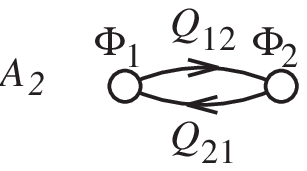}}\nobreak
\medskip\nobreak
\centerline{\fig\figaii{$A_2$ quiver diagram}\ $A_2$ quiver diagram}
\bigskip
For the simplest example, let us study an $A_2$ singularity \DVii. 
The $A_2$ quiver diagram is similar to the $A_2$ Dynkin diagram 
and is represented in \figaii.
We assign a $U(N_i)$ gauge group to the $i$-th node of the $A_2$ diagram.
The dual quiver gauge theory consists of the adjoint scalars $\Phi_1, \Phi_2$ and 
the bifundamental matters $Q_{12}, Q_{21}$ transforming 
in the representation of $(N_1,\bar{N}_2), (N_2,\bar{N}_1)$ respectively. 
The tree level superpotential is 
$$
W(\Phi, Q) = \Tr Q_{12} \Phi_2 Q_{21} - \Tr Q_{21} \Phi_1 Q_{12}
+ W_1(\Phi_1) + W_2(\Phi_2).
$$
The supersymmetry of the quiver gauge theory is broken 
from ${\cal N}=2$ to ${\cal N}=1$ by inserting the superpotentials $W_i(\Phi_i)$.
The partition function of the matrix model which we should consider is 
$Z=\int d\Phi dQ \exp\left(-{1 \over g_s}W(\Phi,Q)\right)$.
$\Phi_i$ is an $N_i \times N_i$ matrix 
and $Q_{ij}$ is an $N_i \times N_j$ matrix.
Integrating $Q_{ij}$ out in the partition function, we obtain 
the effective action of $\Phi_i$.
Since $\Phi_i$ is the $N_i \times N_i$ matrix, 
using the eigen values $\lambda_{i,I}$ $(i=1,2,\quad I=1,\cdots,N_i)$, 
we exchange the matrix integrals $\int d\Phi$ 
for the eigen value integrals $\int d\lambda_{i,I}$.
We then obtain the equations of motion \DVii,
\eqnn\eqmai
\eqnn\eqmaii
$$\eqalignno{
&{1 \over g_s} W'_1(\lambda_{1,I}) 
- 2 \sum_{J = 1, J \neq I}^{N_1} {1 \over \lambda_{1,I} - \lambda_{1,J}}
+ \sum_{J=1}^{N_2} {1 \over \lambda_{1,I} - \lambda_{2,J}}=0, &\eqmai \cr
&{1 \over g_s} W'_2(\lambda_{2,I}) 
- 2 \sum_{J = 1, J \neq I}^{N_2} {1 \over \lambda_{2,I} - \lambda_{2,J}}
+ \sum_{J=1}^{N_1} {1 \over \lambda_{2,I} - \lambda_{1,J}}=0. &\eqmaii
}$$
The second terms  in \eqmai\ and \eqmaii\ are 
the contributions of loop effects of $\lambda_{i,I}$, 
and the third terms come from the contributions of $Q_{ij}$. 
On the analogy of \omv, we define resolvents,
\eqn\reso{
\omega_1(x) \equiv {1 \over N_1} \sum_{I=1}^{N_1} {1 \over x - \lambda_{1,I}}, \quad
\omega_2(x) \equiv {1 \over N_2} \sum_{I=1}^{N_2} {1 \over x - \lambda_{2,I}}.
}

By the way, the classical deformation of the $A_2$ singularity is denoted by 
\eqn\aIIi{
u^2 + v^2 + (y - t_1^{\rm cl}(x))(y - t_2^{\rm cl}(x))(y - t_3^{\rm cl}(x)) = 0,\quad 
\sum_{i=1}^3 t_i^{\rm cl}(x) = 0, 
}
where the deformation parameters $t_i^{\rm cl}(x)$ are given by
\eqn\aIIii{
t_1^{\rm cl}(x) = -{2W'_1(x) + W'_2(x) \over 3}, \quad 
t_2^{\rm cl}(x) = {W'_1(x) - W'_2(x) \over 3}, \quad 
t_3^{\rm cl}(x) = {W'_1(x) + 2W'_2(x) \over 3},
}
from \fibrea\ and \gofan \CIVii.
On the other hand, from \eqmai, \eqmaii\ and \reso, 
we obtain the one-form $y_i(x)dx$ which are described as
\eqn\oneform{
y_1(x) = W'_1(x) - 2 S_1 \omega_1(x) + S_2 \omega_2(x),\quad
y_2(x) = W'_2(x) + S_1 \omega_1(x) - 2 S_2 \omega_2(x).
}
Since the classical deformations are given by 
$y_1^{\rm cl}(x) = t_2^{\rm cl} - t_1^{\rm cl} = W'_1(x)$ and $y_2^{\rm cl}(x) = t_3^{\rm cl} - t_2^{\rm cl} = W'_2(x)$,
the terms including $\omega_i$ in \oneform\ imply quantum effects.
We can define the quantum deformation of the $A_2$ singularity \DVii\ as
\eqn\daIIi{
u^2 + v^2 + (y - t_1(x))(y - t_2(x))(y - t_3(x)) = 0,\quad
\sum_{i=1}^3 t_i(x) = 0, 
}
so that $t_i(x)$ satisfy
$y_1(x) = t_2(x) - t_1(x)$ and $y_2(x) = t_3(x) - t_2(x)$.
Actually $t_i(x)$ can be written down as
\eqn\daIIii{
t_1(x) = t_1^{\rm cl}(x) + S_1 \omega_1(x),\quad 
t_2(x) = t_2^{\rm cl}(x) - S_1 \omega_1(x) + S_2 \omega_2(x), \quad
t_3(x) = t_3^{\rm cl}(x) - S_2 \omega_2(x).
}

So far we have given a brief review on the geometry of the $A_2$ quiver \DVii.
We will generalize the above discussions to the $A_n$ quiver 
and calculate the effective superpotentials of $A_n$ quiver gauge theories.
We will consider, in particular, the quadratic tree level superpotentials.
\bigskip
\centerline{\epsfbox{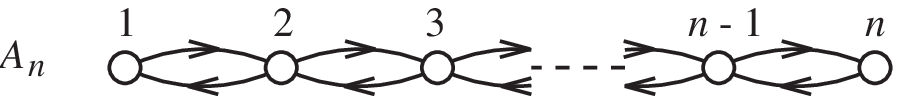}}\nobreak
\medskip\nobreak
\centerline{\fig\figi{$A_n$ quiver diagram}\ $A_n$ quiver diagram}
\bigskip
The $A_n$ quiver diagram is represented in \figi. 
Since the diagram has $n$ nodes, 
we assign a $U(N_i)$ gauge group to each $i$-th node.
The dual theory is the $A_n$ quiver gauge theory 
which consists of the adjoint scalars $\Phi_i$ and 
the bifundamental matters $Q_{i,i+1}, Q_{i+1,i}$.
The tree level superpotential is
$$
W(\Phi,Q) = \sum_{i=1}^{n-1}
\Tr (Q_{i,i+1} \Phi_{i+1} Q_{i+1,i} - Q_{i+1,i} \Phi_i Q_{i,i+1})
+ \sum_{i=1}^{n} \Tr W_i(\Phi_i),
$$
where $W_i(\Phi_i)$ is a polynomial of $\Phi_i$.
We integrate out $Q_{ij}$ in the partition function, 
$$
Z = \int \prod_{i=1}^n d\Phi_i \prod_{i,j}dQ_{ij} \exp\left(-{1 \over g_s}W(\Phi,Q)\right),
$$
and we can rewrite it in terms of the eigen values 
$\lambda_{i,I}$ $(I=1, \cdots, N_i)$ of $\Phi_i$.
We then obtain
\eqnn\anpart
$$
\eqalignno{
Z &= \int \prod_{i=1}^n d\Phi_i 
{1 \over \prod_{i}\det(\Phi_i \otimes {\bf 1} - {\bf 1} \otimes \Phi_{i+1})}
\exp\left(-{1 \over g_s}W(\Phi,Q)\right) & \cr
&= \int \prod_{i,I} d\lambda_{i,I} 
{\prod_{i, I<J} (\lambda_{i,I} - \lambda_{i,J})^2 
\over \prod_{i,I,J}(\lambda_{i,I} - \lambda_{i+1,J})}
\exp\left(-{1 \over g_s}\sum_{i,I}W_i(\lambda_{i,I})\right) . &\anpart
}$$ 
When we describe the partition function 
as $\int \prod_{i,I} d\lambda_{i,I} e^{-\hat{S}}$,
the effective action $\hat{S}$ is denoted by
\eqn\anaction{
\hat{S} = {1 \over g_s} \sum_{i,I} W_i(\lambda_{i,I})
-2 \sum_{i,I<J} \log (\lambda_{i,I}- \lambda_{i,J})
+ \sum_{i,I,J} \log (\lambda_{i,I} - \lambda_{i+1,J}).
}
The equations of motion for the action \anaction\ become
\eqn\aneom{\eqalign{
&{1 \over g_s}W'_1(\lambda_{1,I}) 
- 2\sum_{J = 1,J \neq I}^{N_1}{1 \over \lambda_{1,I} - \lambda_{1,J}}
+ \sum_{J=1}^{N_2}{1 \over \lambda_{1,I} - \lambda_{2,J}} = 0, \cr
&{1 \over g_s}W'_j(\lambda_{j,I}) 
- 2\sum_{J = 1,J \neq I}^{N_j}{1 \over \lambda_{j,I} - \lambda_{j,J}} \cr
&+ \sum_{J=1}^{N_{j-1}}{1 \over \lambda_{j,I} - \lambda_{j-1,J}}
+ \sum_{J=1}^{N_{j+1}}{1 \over \lambda_{j,I} - \lambda_{j+1,J}} = 0, 
\quad j = 2, \cdots, n-1,\cr
&{1 \over g_s}W'_n(\lambda_{n,I}) 
- 2\sum_{J = 1,J \neq I}^{N_n}{1 \over \lambda_{n,I} - \lambda_{n,J}}
+ \sum_{J=1}^{N_{n-1}}{1 \over \lambda_{n,I} - \lambda_{n-1,J}} = 0.
}}
We also introduce the resolvents,
\eqn\resolvent{
\omega_i(x) \equiv {1 \over N_i} \sum_{I=1}^{N_i}{1 \over x - \lambda_{i,I}},
\quad i = 1, \cdots, n,
}
and the 't Hooft couplings $S_i \equiv g_s N_i$.
Note that, in the context of the large $N$ duality, we fix $S_i$ 
and take the limits where $N_i$ go to infinity.
We then read the following functions from \aneom\ as
\eqnn\anofi
\eqnn\anofii
\eqnn\anofiii
$$\eqalignno{
y_1(x) &= W'_1(x) -2 S_1 \omega_1(x) + S_2 \omega_2(x), &\anofi \cr
y_j(x) &= W'_j(x) -2 S_j \omega_j(x) + S_{j-1} \omega_{j-1}(x) 
+ S_{j+1} \omega_{j+1}(x),\quad j = 2, \cdots, n-1, &\anofii \cr
y_n(x) &= W'_n(x) -2 S_n \omega_n(x) + S_{n-1} \omega_{n-1}(x). &\anofiii
}$$
$y_i(x)$ include the quantum deformations of $A_n$ singularity and
$y_i(x)dx$ denote the deformed one-forms.
Note that $y_i^{\rm cl} = W'_i$ are regarded as the classical deformations.

Let us set the tree level superpotentials to be the quadratic ones,
\eqn\sppot{
W_i(x) = {m_i \over 2}x^2,\quad i = 1, \cdots, n.
}
From $W'_i(\lambda_{i,I}) = 0$, the classical vacua are $\lambda_{i,I}=0$.
Since the perturbative analyses around these vacua in the matrix model 
give us the exact effective superpotentials of the ${\cal N}=1$ dual gauge theories, 
we approximate $\lambda_{i,I}$ to the vacuum expectation values, that is, 
we set all $\lambda_{i,I}$ to be equal to zero. 
\anofi\ then becomes
$$
y_1(x) = m_1 x - {2S_1 - S_2 \over x} = {m_1 \over x}(x - a_1^+)(x - a_1^-),
$$
where $a_1^\pm = \pm \sqrt{{2S_1 - S_2 \over m_1}}$.
The critical point $a_1=0$ of $W_1'(a_1)=0$ is splitted to the two points $a_1^\pm$. 
In the same way, each critical point $a_j=0$ $(j=2,\cdots,n-1)$ is
splitted to the points 
$a_j^\pm = \pm \sqrt{{2S_j - S_{j-1} - S_{j+1} \over m_j}}$, 
and $a_n = 0$ is splitted to 
$a_n^\pm = \pm \sqrt{{2S_n - S_{n-1} \over m_n}}$.
In other words, every original critical point is resolved to the two points.
By the use of these resolved points, the periods \omix\ are described as
\eqn\periodint{
\mu_i = {1 \over 2\pi i}\int_{a_i^-}^{a_i^+} y_i(x)dx, \quad 
\Pi_i = \int_{a_i^+}^{\Lambda_i^{3 \over 2}} y_i(x)dx.
}
Since the $B_i$-cycles are non-compact, the cut-off $\Lambda_i$ are needed.
From \anofi, \anofii\ and \anofiii, we obtain the periods around the B-cycles,
\eqnn\anpi
\eqnn\anpii
\eqnn\anpiii
$$\eqalignno{
\Pi_1 =& {1 \over 2}m_1\Lambda_1^3 
- {1 \over 2}(2S_1 - S_2)\left(1 - \log {2S_1 - S_2 \over \Lambda_1^3}\right)
- {1 \over 2}(2S_1 - S_2)\log m_1, &\anpi \cr
\Pi_j =& {1 \over 2}m_j\Lambda_j^3 
- {1 \over 2}(2S_j - S_{j-1} - S_{j+1})\left(1 - \log {2S_j - S_{j-1} - S_{j+1} \over \Lambda_j^3}\right) \cr
&- {1 \over 2}(2S_j - S_{j-1} - S_{j+1})\log m_j,
\quad j=2, \cdots, n-1, &\anpii \cr
\Pi_n =& {1 \over 2}m_n\Lambda_n^3 
- {1 \over 2}(2S_n - S_{n-1})\left(1 - \log {2S_n - S_{n-1} \over \Lambda_n^3}\right) \cr
&- {1 \over 2}(2S_n - S_{n-1})\log m_n. &\anpiii
}$$
We also calculate the periods around the A-cycles as
\eqnn\anmui
\eqnn\anmuii
\eqnn\anmuiii
$$\eqalignno{
\mu_1 =& {1 \over 2}(2S_1 - S_2), &\anmui \cr
\mu_j =& {1 \over 2}(2S_j - S_{j-1} - S_{j+1}),\quad j = 2, \cdots, n-1, &\anmuii \cr
\mu_n =& {1 \over 2}(2S_n - S_{n-1}). &\anmuiii
}$$
Using these results and \introii, we obtain the effective superpotential,
\eqn\aneff{\eqalign{
W_{\rm eff} =& \sum_{i=1}^n {1 \over 2}N_i m_i \Lambda_i^3 \cr
&- {1 \over 2}\bigg[N_1(2S_1 - S_2)\left(1 - \log {2S_1 - S_2 \over \Lambda_1^3}\right) 
+ N_n(2S_n - S_{n-1})\left(1 - \log {2S_n - S_{n-1} \over \Lambda_n^3}\right) \cr
&+ \sum_{i=2}^{n-1} N_i(2S_i - S_{i-1} - S_{i+1})\left(1 - \log {2S_i - S_{i-1} - S_{i+1} \over \Lambda_i^3}\right)\bigg] \cr
&- {1 \over 2}\bigg[N_1(2S_1 - S_2)\log m_1 + N_n(2S_n - S_{n-1})\log m_n \cr
&+ \sum_{i=2}^{n-1}N_i(2S_i - S_{i-1} - S_{i+1})\log m_i\bigg] \cr
&+ \pi i \bigg[\tau_1(2S_1 - S_2) + \tau_n(2S_n - S_{n-1})
+ \sum_{i=2}^{n-1}\tau_i(2S_i - S_{i-1} - S_{i+1})\bigg].
}}
We can reproduce the Veneziano-Yankielowicz terms \VY, which  
appear in the second term of \aneff.
Note that, if we set all $N_i = N$, all $m_i = 1$, all $\Lambda_i = \Lambda$, 
all $S_i = S$ and all $\tau_i = \tau$, the effective superpotential 
is denoted simply by
\eqn\anaffsimp{
W_{\rm eff} = {1 \over 2}nN\Lambda^3 - NS\left(1 - \log {S \over \Lambda^3}\right)
+ 2\pi i \tau S.
}
Note that the constant terms $\sum_{i=1}^n {1 \over 2}N_i m_i \Lambda_i^3$ 
in \aneff\ and ${1 \over 2}nN\Lambda^3$ in \anaffsimp\ can be ignored.

\subsec{$D_n$ singularity}

Next let us consider the $D_n$ singularities. 
A $D_4$ singularity appears in the compactifications of F-theory and
is discussed also in the context of Dijkgraaf-Vafa conjecture \Mc.
\bigskip
\centerline{\epsfbox{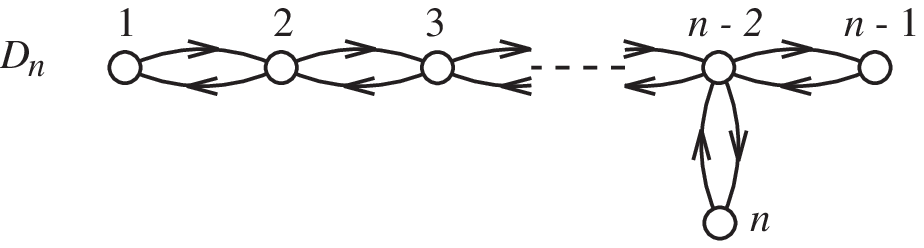}}\nobreak
\medskip\nobreak
\centerline{\fig\figii{$D_n$ quiver diagram}\ $D_n$ quiver diagram}
\bigskip
The $D_n$ quiver diagram is represented in \figii.
In the same way as the $A_n$ quiver gauge theories, 
we assign a $U(N_i)$ gauge group to each $i$-th node.
The fields defined in the $D_n$ quiver gauge theories are
the adjoint scalars $\Phi_i$ for $i=1,\cdots,n$ and 
the bifundamental matters $Q_{ij}$ for the $i$-th and $j$-th nodes 
which are linked to each other. 

The tree level superpotential is
\eqn\dntree{\eqalign{
W(\Phi,Q)=& \sum_{i=1}^{n-1} \Tr (Q_{i,i+1}\Phi_{i+1}Q_{i+1,i} - Q_{i+1,i}\Phi_iQ_{i,i+1}) \cr
&+ \Tr (Q_{n-2,n}\Phi_{n}Q_{n,n-2} - Q_{n,n-2}\Phi_{n-2}Q_{n-2,n})
+ \sum_{i=1}^n \Tr W_i(\Phi_i).
}}
Integrating $Q_{ij}$ out and rewriting the matrix integrals of $\Phi_i$ with
the eigen value integrals of $\lambda_{i,I}$ 
which are the eigen values of $\Phi_i$,
we obtain the equations of motion for $\lambda_{i,I}$. 
From these equations of motion, 
we find the one-forms $y_i(x)dx$ of the deformed $D_n$ singularity,
\eqnn\dnofi
\eqnn\dnofii
\eqnn\dnofiii
\eqnn\dnofiv
\eqnn\dnofv
$$\eqalignno{
y_1(x) =& W'_1(x) -2 S_1 \omega_1(x) + S_2 \omega_2(x), &\dnofi \cr
y_j(x) =& W'_j(x) -2 S_j \omega_j(x) + S_{j-1} \omega_{j-1}(x) 
+ S_{j+1} \omega_{j+1}(x),\quad j=2,\cdots,n-3, &\dnofii\cr
y_{n-2}(x) =& W'_{n-2}(x) -2 S_{n-2} \omega_{n-2}(x) + S_{n-3} \omega_{n-3}(x) 
+ S_{n-1} \omega_{n-1}(x) \cr
&+ S_n \omega_n(x), &\dnofiii\cr
y_{n-1}(x) =& W'_{n-1}(x) -2 S_{n-1} \omega_{n-1}(x) 
+ S_{n-2} \omega_{n-2}(x), &\dnofiv\cr
y_n(x) =& W'_n(x) -2 S_n \omega_n(x) + S_{n-2} \omega_{n-2}(x), &\dnofv
}$$
by the use of the resolvents which are defined in the same way as \resolvent.
In the $D_n$ case different from the $A_n$ case, 
\dnofiii\ is characteristic, because the $(n-2)$-th node is 
linked to the three nodes.
The one-forms $y_i(x)dx$ include the quantum effects 
coming from the $D_n$ matrix models. 

Let us consider the quadratic superpotential \sppot.
We can then calculate the periods around the B-cycles as
\eqnn\dnpi
\eqnn\dnpii
\eqnn\dnpiii
\eqnn\dnpiv
\eqnn\dnpv
$$\eqalignno{
\Pi_1 =& {1 \over 2} m_1\Lambda_1^3 
- {1 \over 2}(2S_1 - S_2)\left(1 - \log {2S_1 - S_2 \over \Lambda_1^3}\right)
- {1 \over 2}(2S_1 - S_2) \log m_1, &\dnpi \cr
\Pi_j =& {1 \over 2} m_j\Lambda_j^3 
- {1 \over 2}(2S_j - S_{j-1} - S_{j+1})\left(1 - \log {2S_j - S_{j-1} - S_{j+1} \over \Lambda_j^3}\right) \cr
&- {1 \over 2}(2S_j - S_{j-1} - S_{j+1}) \log m_j,\quad j=2,\cdots,n-3, &\dnpii \cr
\Pi_{n-2} =& {1 \over 2} m_{n-2}\Lambda_{n-2}^3 
- {1 \over 2}(2S_{n-2} - S_{n-3} - S_{n-1} - S_n)\left(1 - \log {2S_{n-2} - S_{n-3} - S_{n-1} - S_n \over \Lambda_{n-2}^3}\right) \cr
&- {1 \over 2}(2S_{n-2} - S_{n-3} - S_{n-1} - S_n) \log m_{n-2}, &\dnpiii \cr
\Pi_{n-1} =& {1 \over 2} m_{n-1}\Lambda_{n-1}^3 
- {1 \over 2}(2S_{n-1} - S_{n-2})\left(1 - \log {2S_{n-1} - S_{n-2} \over \Lambda_{n-1}^3}\right) \cr
&- {1 \over 2}(2S_{n-1} - S_{n-2}) \log m_{n-1}, &\dnpiv \cr
\Pi_n =& {1 \over 2} m_n\Lambda_n^3 
- {1 \over 2}(2S_n - S_{n-2})\left(1 - \log {2S_n - S_{n-2} \over \Lambda_n^3}\right)
- {1 \over 2}(2S_n - S_{n-2}) \log m_n, &\dnpv
}$$
and the periods around the A-cycles as
\eqnn\dnmui
\eqnn\dnmuii
\eqnn\dnmuiii
\eqnn\dnmuiv
\eqnn\dnmuv
$$\eqalignno{
\mu_1 =& {1 \over 2}(2S_1 - S_2), &\dnmui \cr
\mu_j =& {1 \over 2}(2S_j - S_{j-1} - S_{j+1}),\quad j=2,\cdots,n-3, &\dnmuii \cr
\mu_{n-2} =& {1 \over 2}(2S_{n-2} - S_{n-3} - S_{n-1} - S_n), &\dnmuiii \cr
\mu_{n-1} =& {1 \over 2}(2S_{n-1} - S_{n-2}), &\dnmuiv \cr
\mu_n =& {1 \over 2}(2S_n - S_{n-2}). &\dnmuv
}$$
From these periods, we can obtain the effective superpotentials
in terms of \introii.

For example, we consider the $D_4$ quiver gauge theory.
The effective superpotential becomes
\eqn\divsuperpot{\eqalign{
W_{\rm eff} =& 
- {1 \over 2}N(2S_1 - S_2)\left(1 - \log {2S_1 - S_2 \over \Lambda^3}\right)
- {1 \over 2}N(2S_3 - S_2)\left(1 - \log {2S_3 - S_2 \over \Lambda^3}\right) \cr
&- {1 \over 2}N(2S_4 - S_2)\left(1 - \log {2S_4 - S_2 \over \Lambda^3}\right) \cr
&- {1 \over 2}N(2S_2 - S_1 - S_3 - S_4)\left(1 - \log {2S_2 -2 S_1 - S_3 - S_4 \over \Lambda^3}\right) \cr
& + \pi i \tau (S_1 + S_3 + S_4 - S_2),
}}
where, for simplicity, we set that all $m_i = 1$, all $N_i = N$, 
all $\Lambda_i=\Lambda$ and all $\tau_i = \tau$, 
and the constant term $\sum_i{1 \over 2}N_im_i\Lambda_i^3$ is ignored.
Since the first, third and fourth nodes of the $D_4$ quiver diagram 
have a cyclic symmetry, $S_1, S_3, S_4$ in
the superpotential \divsuperpot\ can be replaced with one another.

\subsec{$E_n$ singularity}

Finally we consider the $E_n$ singularities. 
In the string theories, the $E_n$ singularities play important roles.
For example, $E_8 \times E_8$ gauge symmetry of heterotic string theories
are realized by the $E_n$ singular fibres in the F-theory.
So it is interesting to analyse the $E_n$ singularities.
\bigskip
\centerline{\epsfbox{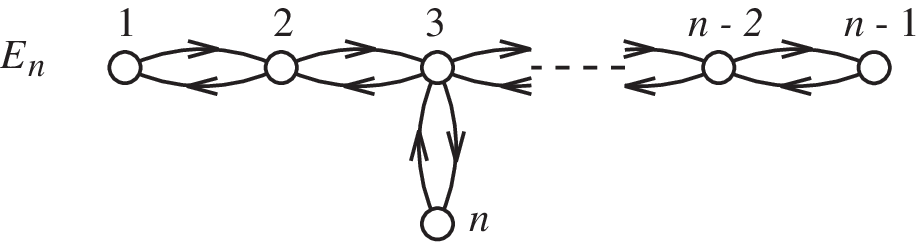}}\nobreak
\medskip\nobreak
\centerline{\fig\figiii{$E_n$ quiver diagram}\ $E_n$ quiver diagram}
\bigskip
The $E_n$ quiver diagram is depicted in \figiii. 
Each $i$-th node has a $U(N_i)$ gauge group.
In the same way as the $A_n$ and $D_n$ quiver gauge theories, 
we define the adjoint scalars $\Phi_i$ and the bifundamental matters $Q_{ij}$.
The tree level superpotential in the $E_n$ quiver matrix models is
\eqn\entree{\eqalign{
W(\Phi,Q)=&
\sum_{i=1}^{n-1}  \Tr(Q_{i,i+1}\Phi_{i+1}Q_{i+1,i} - Q_{i+1,i}\Phi_iQ_{i,i+1}) \cr
&+ \Tr (Q_{3n}\Phi_nQ_{n3} - Q_{n3}\Phi_3Q_{3n}) 
+ \sum_{i=1}^n \Tr W_i(\Phi_i).
}}
We integrate $Q_{ij}$ out in the partition function 
$Z=\int d\Phi dQ \exp \left(-{1 \over g_s} W(\Phi,Q)\right)$ and
obtain the effective action of $\lambda_{i,I}$ 
which are the eigen values of $\Phi_i$.
We calculate the equations of motion from this effective action 
and read the following functions,
\eqnn\enofi
\eqnn\enofii
\eqnn\enofiii
\eqnn\enofiv
\eqnn\enofv
\eqnn\enofvi
$$\eqalignno{
y_1(x) =& W'_1(x) - 2S_1 \omega_1(x) + S_2 \omega_2(x), &\enofi \cr
y_2(x) =& W'_2(x) - 2S_2 \omega_2(x) + S_1 \omega_1(x) + S_3 \omega_3(x), &\enofii \cr
y_3(x) =& W'_3(x) - 2S_3 \omega_3(x) + S_2 \omega_2(x) + S_4 \omega_4(x) + S_n \omega_n(x), &\enofiii \cr
y_j(x) =& W'_j(x) - 2S_j \omega_j (x)+ S_{j-1} \omega_{j-1}(x)  \cr
&+ S_{j+1} \omega_{j+1}(x),\quad j=4,\cdots,n-2, &\enofiv \cr
y_{n-1}(x) =& W'_{n-1}(x) - 2S_{n-1} \omega_{n-1}(x) + S_{n-2} \omega_{n-2}(x), &\enofv \cr
y_n(x) =& W'_n(x) - 2S_n \omega_n(x) + S_3 \omega_3(x), &\enofvi
}$$
We also consider the quadratic superpotential \sppot\ and
assume the eigen values in $\omega_i(x)$ to be in 
the vacua, that is, $\lambda_{i,I}=0$.
Since every critical point of $W_i(x)$ is then splitted to the two points
by the quantum deformations,  
we calculate the period \periodint\ for the one-form $y_i(x)dx$.
The periods $\Pi_i$ around the B-cycles are
\eqnn\enpi
\eqnn\enpii
\eqnn\enpiii
\eqnn\enpiv
\eqnn\enpv
\eqnn\enpvi
$$\eqalignno{
\Pi_1 =& {1 \over 2}m_1\Lambda_1^3
-{1 \over 2}(2S_1 - S_2)\left(1-\log{2S_1 - S_2 \over \Lambda_1^3}\right)
-{1 \over 2}(2S_1 - S_2) \log m_1, &\enpi \cr
\Pi_2 =& {1 \over 2}m_2\Lambda_2^3
-{1 \over 2}(2S_2 - S_1 - S_3)\left(1-\log{2S_2 - S_1 - S_3 \over \Lambda_2^3}\right) \cr
&-{1 \over 2}(2S_2 - S_1 - S_3) \log m_2, &\enpii \cr
\Pi_3 =& {1 \over 2}m_3\Lambda_3^3
-{1 \over 2}(2S_3 - S_2 - S_4 - S_n)\left(1-\log{2S_3 - S_2 - S_4 - S_n \over \Lambda_3^3}\right) \cr
&-{1 \over 2}(2S_3 - S_2 - S_4 - S_n) \log m_3, &\enpiii \cr
\Pi_j =& {1 \over 2}m_j\Lambda_j^3
-{1 \over 2}(2S_j - S_{j-1} - S_{j+1})\left(1-\log{2S_j - S_{j-1} - S_{j+1} \over \Lambda_j^3}\right) \cr
&-{1 \over 2}(2S_j - S_{j-1} - S_{j+1}) \log m_j,\quad j=4,\cdots,n-2, &\enpiv \cr
\Pi_{n-1} =& {1 \over 2}m_{n-1}\Lambda_{n-1}^3
-{1 \over 2}(2S_{n-1} - S_{n-2})\left(1-\log{2S_{n-1} - S_{n-2} \over \Lambda_{n-1}^3}\right) \cr
&-{1 \over 2}(2S_{n-1} - S_{n-2}) \log m_{n-1}, &\enpv \cr
\Pi_n =& {1 \over 2}m_n\Lambda_n^3
-{1 \over 2}(2S_n - S_3)\left(1-\log{2S_n - S_3 \over \Lambda_n^3}\right)
-{1 \over 2}(2S_n - S_3) \log m_n. &\enpvi
}$$
and the periods $\mu_i$ around the A-cycles are
\eqnn\enmui
\eqnn\enmuii
\eqnn\enmuiii
\eqnn\enmuiv
\eqnn\enmuv
\eqnn\enmuvi
$$\eqalignno{
\mu_1 =& {1 \over 2}(2S_1 - S_2), &\enmui \cr
\mu_2 =& {1 \over 2}(2S_2 - S_1 - S_3), &\enmuii \cr
\mu_3 =& {1 \over 2}(2S_3 - S_2 - S_4 - S_n), &\enmuiii \cr
\mu_j =& {1 \over 2}(2S_j - S_{j-1} - S_{j+1}),\quad j=4,\cdots,n-2, &\enmuiv \cr
\mu_{n-1} =& {1 \over 2}(2S_{n-1} - S_{n-2}), &\enmuv \cr
\mu_n =& {1 \over 2}(2S_n - S_3). &\enmuvi
}$$
From these periods and \introii, we can calculate 
the effective superpotentials of the $E_n$ quiver gauge theories.
For simplicity, we ignore the terms independent of $S_i$ and 
set that all $m_i = 1$, all $N_i = N$, 
all $\Lambda_i=\Lambda$ and all $\tau_i = \tau$.
For example the effective superpotential of the $E_8$ quiver then becomes
$$\eqalign{
W_{\rm eff} =& 
-{1 \over 2}N(2S_1 - S_2)\left(1-\log{2S_1 - S_2 \over \Lambda^3}\right)
-{1 \over 2}N(2S_7 - S_6)\left(1-\log{2S_7 - S_6 \over \Lambda^3}\right) \cr
&-{1 \over 2}N(2S_8 - S_3)\left(1-\log{2S_8 - S_3 \over \Lambda^3}\right) \cr
&-{1 \over 2}N(2S_3 - S_2 - S_4 - S_8)\left(1-\log{2S_3 - S_2 - S_4 - S_8 \over \Lambda^3}\right) \cr
&-{1 \over 2}N\sum_{j=2,4,5,6}(2S_j - S_{j-1} - S_{j+1})\left(1-\log{2S_j - S_{j-1} - S_{j+1} \over \Lambda^3}\right) \cr
&+\pi i \tau (S_1 + S_7 + S_8 - S_3).
}$$
In this effective superpotential we can also find Veniziano-Yankielowicz terms.
Since the third node is linked to the three nodes, 
$S_3$ is characteristic in the $E_n$ quiver as well as in the $D_n$ quiver.

\newsec{Conclusions and discussions}

We have considered the Calabi-Yau manifolds with the ADE singularities.
If we calculate the periods of one-forms $y_i(x)dx$ 
around compact A-cycles and non-compact B-cycles
on the deformed geometry of the Calabi-Yau manifolds,
we can obtain the effective superpotentials of the dual gauge theories 
by the geometric engineering. 
We have calculated the equations of motion in the ADE multi-matrix models.
Since the quantum deformations are derived from the 
perturbative analyses of the matrix models by the Dijkgraaf-Vafa conjecture, 
We have found out the quantum deformations of the one-forms $y_i(x)dx$ 
from those equations of motion in the ADE matrix models.

We have considered the quadratic superpotentials 
$W_i(\Phi_i) = {1 \over 2} m_i \Tr \Phi_i^2$ 
for the simple examples. Then the classical vacua are $\lambda_{i,I}=0$, 
where $\lambda_{i,I}$ are the eigen values of $\Phi_i$.
Since the perturbation theory on these vacua gives rise to 
the effective superpotential in the dual gauge theory, 
we have approximated $\lambda_{i,I}$ in the resolvents $\omega_i$ 
to the vacua. The original critical point $a_i=0$, which is obtained 
from $(y_i^{\rm cl} =) W_i'(a_i) = 0$, is splitted to the 
two points $a_i^\pm$, which are derived 
from $y_i(a_i^\pm) = 0$ on the deformed geometry.
In terms of $a_i^\pm$ and the cut-off parameters $\Lambda_i$,
we have calculated the periods \periodint.
From these periods we have written down the effective superpotentials of 
the dual quiver gauge theories. 
We have also found that the effective superpotentials include 
the Veneziano-Yankielowicz terms.

We have used the approximation, that is, all eigen values 
of $\Phi_i$ appearing in the resolvents are in the classical vacua.
But in order to derive exact effective superpotentials for 
the ${\cal N}=1$ quiver gauge theories, we should achieve 
the integration of the eigen values in the multi-matrix model partition function.

Though it is difficult to exactly calculate the partition functions of 
the multi-matrix models,
we can analyse the loop expansions of the planar diagrams
order by order of the 't Hooft couplings $S_i$ by the use of Feynman diagrams \DGKV. 
So we should confirm the expansions in the 
context of the geometric engineering.

\bigbreak\bigskip\bigskip\centerline{{\bf Acknowledgement}}\nobreak

I would like to thank B. Taylor for useful comments.
This work is supported in part by the Grant-in-Aid for Scientific 
Research of Professor H. Sonoda, Kobe Univ., from Japan Ministry 
of Education, Science and Culture (\#14340077).
\listrefs

\bye